\documentclass[12pt]{iopart}
\usepackage{graphicx}
\usepackage{times,mathptm} 
\usepackage[T1]{fontenc}
\begin{document}
\jl{21}
\title[Microwave properties of FIB weak link]{The microwave power handling of a FIB generated weak
link in a YBCO film}
\author  {A Cowie\dag, L F Cohen\dag, and  M W Denhoff\ddag}
\address{\dag\ Blackett Laboratory, Imperial College, Prince Consort Rd, London, SW7 2BZ }
\address{\ddag\ Institute for Microstructural Sciences, National Research Council, Canada}

\begin{abstract}
We have measured the power dependent microwave properties of a weak link 
in a YBa$_2$Cu$_3$O$_{7-\delta}$ thin film formed by writing a line of damage
using a focused ion beam. The measurement was made using a parallel plate
resonator at 5.5~GHz with the weak link written across the width of one of the plates.
The ion induced damage was characterized using a TRIM computer simulation and
the dc properties of similar weak links was measured. Using a 200 eV Si ion dose of 
$2\times10^{13}\ {\textrm{cm}}^{-2}$, the $T_c$ of the damaged region was reduced by
5.5~K and the normal resistivity was doubled. Surprisingly, the microwave measurements
did not show any Josephson junction characteristics. Rather, the ion damaged region 
exhibited a greatly increased microwave resistivity that was constant as a function 
of microwave power up to rf fields of 20 mT at 21~K.
\end{abstract}


\noindent {\emph{Revised}}: March 26, 1999

\submitted
\maketitle

\section{Introduction} The presence of weak-links in
granular material can give rise to diverse dependence of surface
resistance upon microwave magnetic field. In this context, weak links
are usually modeled in terms of the resistively shunted junction (RSJ)
model~\cite{hylton,attanasio,portis91,portis92,nguyen,oates93}. 
Oates \emph{et al} performed a control experiment using a stripline
resonator geometry with an edge junction and found that the weak link
followed an RSJ type behaviour~\cite{oates96}. In the
present study we used a  parallel plate resonator
geometry with a macroscopic weak link formed by focused ion beam
(FIB) irradiation across one of the films. There are significant
differences between edge junctions and FIB weak links and it is interesting to test 
whether an FIB weak link will also follow RSJ behaviour.

Xie \emph{et al}~\cite{xie} developed an RSJ based model to describe nonlinear
absorption of microwaves in a weak-link Josephson junction. This
predicts that for an ideal Josephson junction, the microwave power
absorbed (power absorbed is proportional to the surface resistance) will
remain independent of the microwave screening current up to a threshold
value dictated by the critical current density of the junction,
$J_{c}$. Further sharp steps in microwave absorption occur periodically
with increasing microwave current as a result of dynamic flux
quantization where one or more fluxons pass through the junction during
each microwave cycle. 

We expect to observe a significant deterioration in surface resistance for 
microwave currents exceeding  $J_{c}$  and further increase in the
surface impedance as microwave currents grow larger. 
(At 77~K, $J_{c}\sim 3\times10^4\ \textrm{A}\!\!\cdot\!\textrm{cm}^{-2}$, which for our films would give a 
microwave field of $B_{rf}\sim 0.08\ \textrm{m}\!\cdot\!\!\textrm{T}$.) 
The long FIB pattern would act as a parallel network of
sub-junctions rather than a single junction, so that the detail of flux
penetration steps would be averaged out in a network of such
sub-junctions~\cite{herd}.

\section{Experimental Method} 
Three YBa$_2$Cu$_3$O$_{7-\delta}$ thin films on CeO$_2$ buffered, sapphire substrates, 
grown by pulsed laser deposition, were supplied by the Institute for Microstructural Sciences of the
National Research Council of Canada. 
The growth method of these films is described in Ref.~\cite{denhoff1}.
The films were 150~nm thick at the
edges and 190~nm thick at the center. Two virgin films, V$_1$ and V$_2$ were
used as a reference set for the parallel plate resonator. The third film
J$_1$, had a bisecting 50~nm line of ion damage written across its full
width by FIB. V vs I dc measurements  on similar weak links have been reported
previously~\cite{denhoff}.
A pair of YBa$_2$Cu$_3$O$_{7-\delta}$ films on LaAlO$_3$ substrates from DuPont, V$_3$ and V$_4$, 
were also measured for further reference.

Microwave measurements were made using a parallel plate resonator consisting of two
10~mm by 5~mm HTS films, separated by a 0.15~mm thick, high purity sapphire
spacer. This structure has a fundamental resonance frequency at
5.5~GHz. The losses are associated with the effective surface
resistance $R_s$ (not corrected for finite sample thickness), dielectric
and radiation losses. This can be expressed as~\cite{taber}
\[ Q^{-1} = \tan \delta + \alpha s + \beta \left( \frac{R_s}{s}\right) . \]
The first two terms are due to loss in the dielectric and radiation loss.
$\tan \delta$  
represents the dielectric loss in the spacer between the two superconducting films. 
$\alpha s$ gives the radiation loss.  These are small
for the geometry and materials used here and will be neglected~\cite{ghosh}.
The term 
$\beta ( {R_s}/{s})$ is the loss due to the surface resistance of the two superconducting 
films. 
 Using $\beta = 1/(\pi \mu_0 f)$, the surface resistance can then be calculated directly from the measured $Q$.

Overall, the microwave
measurement has a precision of 5~$\mu\Omega$ and the signal sensitivity is lost for
$R_s$ values greater than 20~m$\Omega$ . The microwave measurements were performed
in a continuous flow cryostat with a temperature stability within 0.2~K.
The TE10 mode was used. This mode has maximum screening current in the
central band of the film coincident with the FIB line. 

\section{Fabrication of a Focused Ion Beam Junction} 
\begin{figure}[tb]
\centering  
\includegraphics[width=10cm]{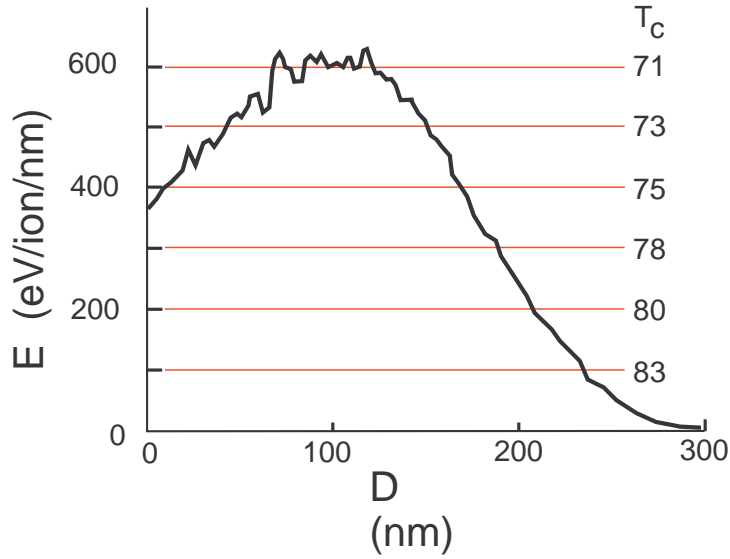}
\caption{Calculated non-ionizing energy loss for 200 keV Si ions incident on YBa$_2$Cu$_3$O$_{7-\delta}$. The $T_c$ 
values are our estimate for a dose of $2\times10^{13}$~cm$^{-2}$, starting with a $T_c$ 
for the undamaged YBa$_2$Cu$_3$O$_{7-\delta}$ of 85~K.}
\label{trimybc2}
\end{figure}

One method of
producing a weak link is to locally damage the superconducting
material using a focused ion beam (FIB). In our case, we used a 200~keV Si ion beam focused 
to a spot size of 50~nm. It has been reported that the $T_c$ reduction due to ion irradiation is 
proportional to the non-ionizing energy loss~\cite{tinchev,summers}.
TRIM~\cite{ziegler} simulation showed that the damage due to the 200~keV Si ions incident on a YBa$_2$Cu$_3$O$_{7-\delta}$
thin film varies with distance into the film, shown in Fig.~\ref{trimybc2}.
The reduction of $T_c$ will depend on depth and will be scaled by the ion dose.
The TRIM simulation also shows that the beam spreads laterally and that the total width of 
the damaged region will be about 200 nm.

\begin{figure}[tb]
\centering  
\includegraphics[width=10cm]{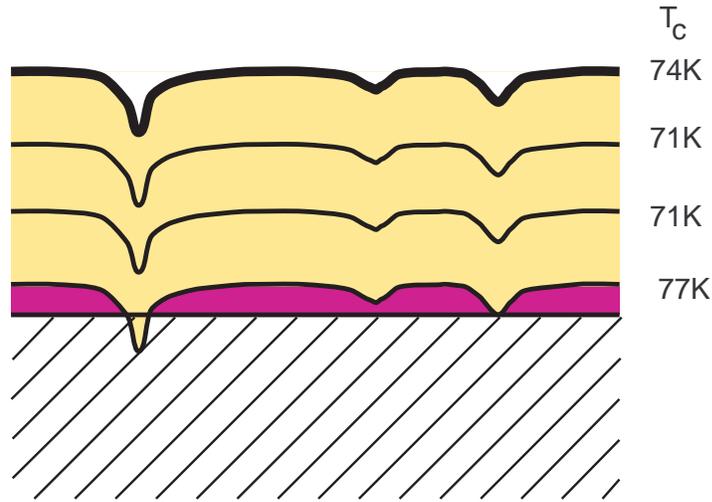}
\caption{Model of the cross-section of an ion damaged YBa$_2$Cu$_3$O$_{7-\delta}$ film, showing lines of constant
$T_c$. At 77~K only the bottom (dark) layer of YBa$_2$Cu$_3$O$_{7-\delta}$ would be superconducting.}
\label{section}
\end{figure}

Since a typical YBa$_2$Cu$_3$O$_{7-\delta}$ film is not smooth, the reduction of $T_c$ will tend to follow the surface 
profile. Figure~\ref{section} shows a model of a cross-section with lines of constant $T_c$. At 
a temperature of 77~K only the bottom layer of YBa$_2$Cu$_3$O$_{7-\delta}$ will be superconducting. The various 
valleys in the YBa$_2$Cu$_3$O$_{7-\delta}$ film (which occur at grain boundaries) would give rise to a variety of 
types of weak links. A large valley would yield an SNS structure and could give rise to a 
Josephson junction based on the proximity effect. A shallow valley would have a thin 
superconducting region next to the substrate and would be a flux flow type of weak link. In any case, 
the bulk of the YBa$_2$Cu$_3$O$_{7-\delta}$ of the top portions of the film will act as a normal shunt of the weak 
links.  At temperatures below 71~K, the entire thickness of the YBa$_2$Cu$_3$O$_{7-\delta}$ film will be 
superconducting.

\section{dc characterization}
\begin{figure}[tb]
\centering  
\includegraphics[width=10cm]{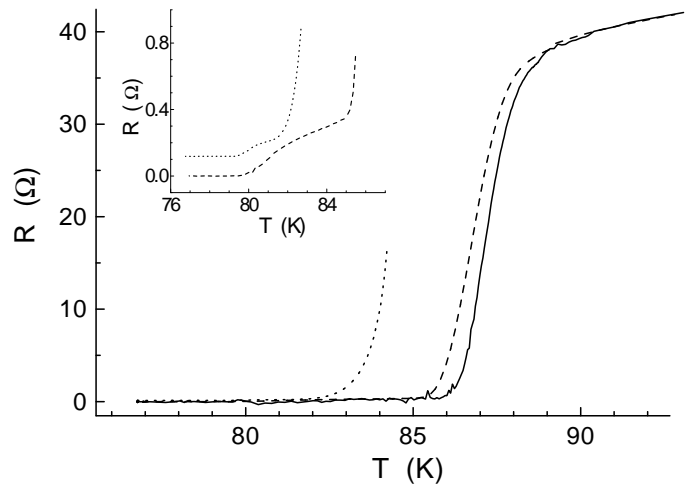}
\caption{Resistance vs temperature data for a 10 $\mu$m wide channel of a YBa$_2$Cu$_3$O$_{7-\delta}$ film written 
with a single FIB line. The measuring 
currents were: full - $10^{-6}$ A, dashed - $10^{-4}$ A, and dotted $10^{-2}$ A. The inset is a blow-up
of the foot of the transition.}
\label{l9615rt}
\end{figure}

A YBa$_2$Cu$_3$O$_{7-\delta}$ film similar to the one we used to make microwave measurements, was patterned into 
four point probe structures to allow dc measurements. The results for a 10 $\mu$m wide 
channel with a single FIB line written across it at a dose of  $2\times10^{13}$~cm$^{-2}$ are 
given in Figure~\ref{l9615rt}.
The main transition shows that the undamaged YBa$_2$Cu$_3$O$_{7-\delta}$ has a $T_c$ of 86~K. A detailed view of the 
small resistance part of the data, reveals a small transition due to the ion damaged line 
with a $T_c$ of 80.5~K for very small currents. This is the temperature where the first 
continuous  superconducting path forms. The critical current of the weak link, at 77~K, was 0.6 mA, 
which is 25 to 50 times less 
than it would be for an undamaged film. From the small transition, we find that the  normal 
resistance of the damaged 
line is 0.34 $\Omega$ at 85~K.
\begin{figure}[b]
\centering  
\includegraphics[width=10cm]{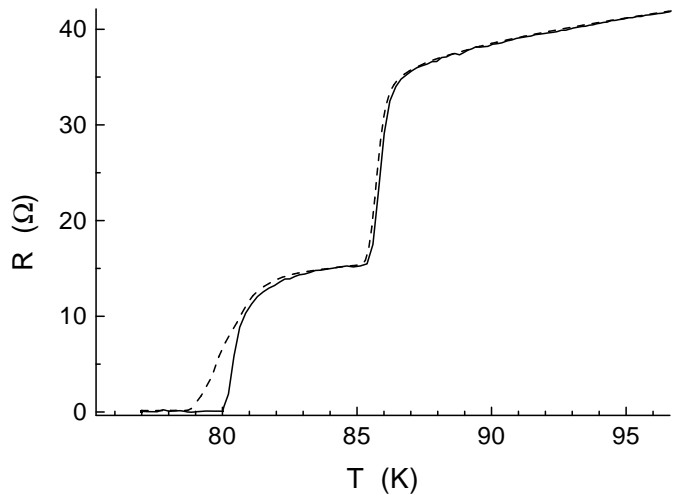}
\caption{Resistance vs temperature data for a 20 $\mu$m wide channel of a YBa$_2$Cu$_3$O$_{7-\delta}$ film written 
with a 20 $\mu$m by 20 $\mu$m FIB area. The measuring 
currents were: full - $10^{-6}$ A and dashed - $10^{-4}$ A.}
\label{l9615rt2}
\end{figure}

In order to further investigate the properties of the ion damaged material, a 20~$\mu$m by 
20~$\mu$m square was damaged using the focused ion beam with the same dose of 
$2\times10^{13}$~cm$^{-2}$. As can be seen in Figure~\ref{l9615rt2}, the transition of the 
damaged area is now a large feature. $T_c$ is 80.5~K, the same as for the single FIB line, 
which implies that the ion damage is the same for a single line and a large area. 
Since the size of the damaged region is well defined, the measured resistance of 15 $\Omega$ 
gives a value of the resistivity of 230 $\mu\Omega\cdot$cm at 85~K. This is about double the 
measured resistivity of the undamaged YBa$_2$Cu$_3$O$_{7-\delta}$ which, extrapolated down to 85~K, was 114 
$\mu\Omega\cdot$cm. Using this resistivity measurement and the resistance measurement of the 
single FIB written line, the effective width of the ion damaged region can be found. It is 
200 nm, which is in agreement with the value predicted by TRIM.
\section{Microwave results} 

\begin{table}[b]
\caption{Low power characteristics of films used in this study.}\label{table:results}
\begin{indented}
\item[]\begin{tabular}{@{}lllll}
\br
Film    & T$_c$& Treatment & R$_s$, 21 K    & R$_s$, 75 K \\
\mr
V$_1$   & 87 K& as-grown & ~100 $\mu\Omega$ & ~390 $\mu\Omega$ \\
V$_2$   & 87 K & as-grown & ~40 $\mu\Omega$ & ~160 $\mu\Omega$ \\
V$_3$   & 90 K & as-grown & ~50 $\mu\Omega$ & ~140 $\mu\Omega$ \\
V$_4$   & 90 K & as-grown & ~40 $\mu\Omega$ & ~140 $\mu\Omega$ \\
J$_1$   & 84.5 K& FIB track  & ~60 $\mu\Omega$ & ~350 $\mu\Omega$ \\
\br
\end{tabular}
\end{indented}
\end{table}

Various pairs of the YBa$_2$Cu$_3$O$_{7-\delta}$ films were measured with the parallel plate
resonator. A summary of the films used in this study is given in Table~\ref{table:results}. V$_1$,
V$_2$ and J$_1$ were supplied by the National Research Council of Canada while
V$_3$ and V$_4$ are YBa$_2$Cu$_3$O$_{7-\delta}$ films from DuPont. 

\begin{figure}[tb]
\centering  
\includegraphics[width=10cm,clip]{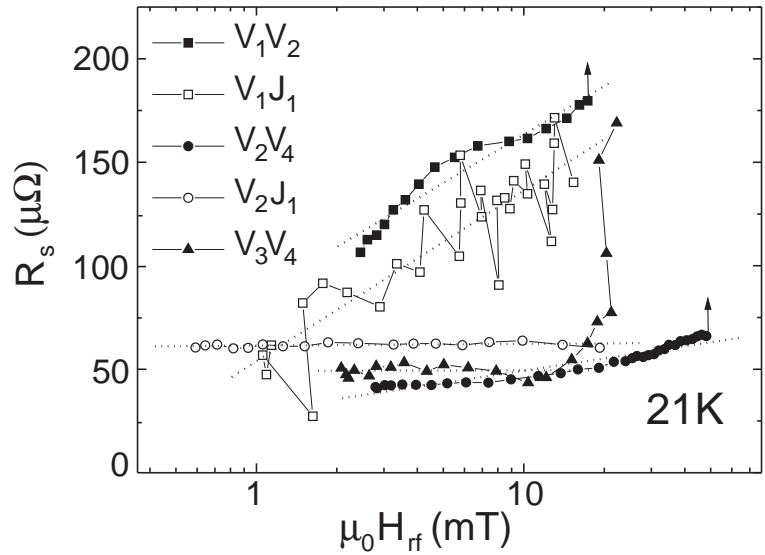}
\caption{Microwave field dependence of the effective surface resistance of
film pairs as labeled, at 21~K.}
\label{ofig3}
\end{figure}

Figure~\ref{ofig3} shows the   
effective surface resistance behaviour for a number of combinations of films at 21~K.
The combination of V$_2$V$_4$
has the lowest residual value of $R_s$ and shows catastrophic failure at
50~mT, but is nonlinear with an increase in $R_s$ of 35 $\mu\Omega$ from 2--50~mT. The
V$_3$V$_4$ combination is independent of power up to 8~mT, shows a region of
increasing nonlinearity and fails catastrophically at 20~mT. V$_4$ is
clearly a ``cleaner'' film than V$_3$. From the nature of the failure of V$_3$V$_4$
and the absolute value of $R_s$ it is probable that there is one defective
region close to the edge of the film V$_3$. 

The V$_2$J$_1$ combination also
displays extremely good power handling at 21~K. The absolute value of $R_s$
is higher in J$_1$ than in V$_2$, V$_3$, or V$_4$ but there is no evidence of
nonlinear behaviour up to 20~mT. While the V$_1$V$_2$ combination shows that V$_1$
is the worst film of the batch. Quasi logarithmic power
dependence is observed at all powers and this parallel plate film
combination fails catastrophically at 20~mT. At 21~K the V$_1$J$_1$ combination
principly reflects the granular behaviour of V$_1$, although the data is
much noisier for this set, the average value given by the dashed line
has a lower overall $R_s$ value than for the virgin film combination. The
noise in the data arose from poor temperature stability ($\pm$ 0.2~K)
during this particular run. 

These results indicate that V$_1$ is a virgin
film with some other microstructural features which provide a mechanism
for power dependence. J$_1$ has been damaged by FIB but this does not seem
to have changed it's power dependence however, its residual surface
resistance has clearly been increased by this process. 

\begin{figure}[tb]
\centering  
\includegraphics[width=10cm,clip]{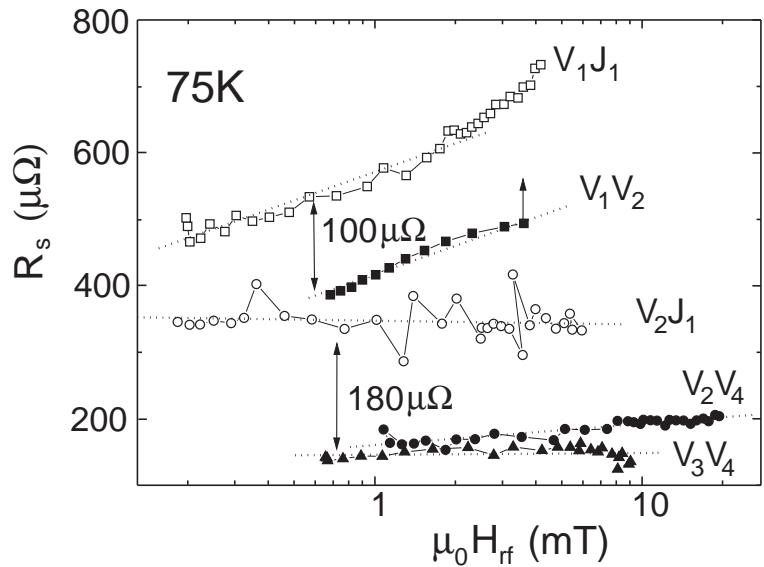}
\caption{Microwave field dependence of the effective surface resistance of
film pairs as labeled, at 75~K.}
\label{ofig4}
\end{figure}

Figure~\ref{ofig4} shows 
the effective surface resistance at 75~K. The V$_3$V$_4$ and V$_2$V$_4$ film
combinations are almost indistinguishable. They display similar power
dependence up until 10~mT where V$_3$V$_4$ fails. Their residual surface
resistances are also very similar. The V$_2$J$_1$ combination shows that J$_1$
does not change the power dependence of this combination overall but
does increase the residual surface resistance. 

The poor microstructure
of V$_1$ dominates the power dependence of the V$_1$V$_2$ and V$_1$J$_1$ combinations
showing logarithmic power dependence in both cases. Again, the influence
of the FIB line upon the behaviour of J$_1$ is to raise it's residual
surface resistance. At 75~K the difference in $R_s$ between V$_2$ and J$_1$ is
about 100~$\mu\Omega$. 

Measurement became difficult above 75~K when the microwave
response is limited by background noise. Neither at 21~K or at 75~K is
there any evidence for junction-type behaviour in parallel plate
combinations using the FIB film J$_1$.

\section{Analysis}
In this section, we will concentrate on the effect of the FIB
line. The measurements on the pair V$_2$J$_1$ yield a combined
"apparent" surface resistance of 60 $\mu\Omega$ (at 21~K), which is
independent of microwave power. Measurements on the films V$_2$, V$_3$,
and V$_4$ show that the surface resistance of V$_2$ is about 40
$\mu\Omega$. The extra loss seen by the measurement must be due
to film J$_1$. There are two possible sources of the loss; 1) the
surface resistance of the area of the film and 2) the FIB damaged
line. 

The microwave properties of J$_1$ were not measured before the FIB
line was written, so we cannot be sure of the source of the extra
loss. However, if the loss is due to a poor quality YBa$_2$Cu$_3$O$_{7-\delta}$ film, we
would expect to see a strong power dependence as was shown by
sample V$_1$. Our dc measurements on other FIB damaged samples,
showed that $T_c$
and $J_c$ were reduced and that the normal dc resistivity was doubled. From
this we expect that the FIB damage should also have an effect on
the microwave properties. We conclude that it is likely that, at
least some of, the extra microwave loss is due to the FIB damaged
line.

In order to be more quantitative in the analysis of the data, a
term representing the FIB line can be added to the parallel plate
resonator model [11]. We will assume that the $R_s$ values of the two films
are the same and that there is a line resistor across only one of
the films. The extended model can be written, 
\[ Q^{-1} = \tan \delta + \alpha s + \beta \left( \frac{R_s}{s}\right) + %
            \gamma \left(\frac{R_l}{s L}\right). \]
The first three terms on the right hand side are the usual ones. 
The last term represents the FIB resistor.  The FIB line is represented by
a linear resistivity  which has units of $R_l =[\Omega\!\cdot\!\textrm{m}]$. 
$L$ is the length of the plate in the direction of the microwave current.
The FIB line was written across the width of the substrate and its length is $W$.
For a uniform current flowing in the $L$ direction, the total resistance
  is $R ={R_l}/{W}$.  $\gamma$ is the geometric factor, which can be 
found by solving the electromagnetic problem for the geometry of the parallel plate 
resonator. For the specific microwave mode used in this experiment, the 
  calculation gives $\gamma = {1}/{\pi \mu_0 f}$.
  
If we assume that the difference between the microwave losses for the pair of films 
V$_2$V$_4$ and the pair V$_2$J$_1$ is only due to the FIB damaged line, then a value of $R_l$ 
can be found. 
Due to this assumption, the deduced resistances should be considered as maximum values.
We find that at 21~K  $R_l = 2\times10^{-7}\ \Omega\cdot$m and at 75~K $R_l = 
2\times10^{-6}\ \Omega\cdot$m.  The lumped element resistance value can be found by dividing 
by the length of the FIB line. This gives $4\times10^{-5}\ \Omega$ at 21~K and 
$4\times10^{-4}\ \Omega$ at 75~K. In order to compare these results with more familiar 
materials, these values can be converted to equivalent volume resistivity in the standard way by 
multiplying by the cross-sectional area and dividing by the length of the current path. This 
gives effective resistivities, at 5.5 GHz, of $\rho_{eff} = 1.7\times10^{-7}\ \Omega\cdot$m 
at 21~K and  $\rho_{eff} = 1.7\times10^{-6}\ \Omega\cdot$m at 75~K.

An effective resistivity can be found from the measured $R_s$ of our undamaged YBa$_2$Cu$_3$O$_{7-\delta}$, which is 
$4\times10^{-5}\ \Omega$ at 21~K. Multiplying this by the film thickness gives 
$ \rho_{eff}  = 7.2 \times 10^{-12}\ \Omega\!\cdot\!\textrm{m}$. 
Similarly, at 75~K, $\rho_{eff} = 2.5 \times 10^{-11}\ 
\Omega\!\cdot\!\textrm{m}$.
The dc resistivity of pure copper is $\rho = 2 \times 10^{-8}\ \Omega\!\cdot\!\textrm{m}$ at 300~K,
 about  $\rho = 2 \times 10^{-9}\ \Omega\!\cdot\!\textrm{m}$ at 77 K, and about
 $\rho = 1 \times 10^{-10}\ \Omega\!\cdot\!\textrm{m}$ at 21 K. 
For comparison, the surface resistances calculated from these resistivities are, 
$R_s = 2.1\times10^{-2}~\Omega$ at 300~K, $R_s = 6.6\times10^{-3}~\Omega$ at 77~K,
and $R_s = 1.5\times10^{-3}~\Omega$ at 21~K.
In the case of our line resistor (area
$= (5\times10^{-3}~\textrm{m}) \times (1.8\times10^{-7}\textrm{m})$ and current path length 
$= 2\times10^{-7}~\textrm{m}$), lumped element resistances can be calculated. 
These values are  summarized in table~\ref{table:rho}.
\begin{table}
\caption{Equivalent lumped element resistance for a line resistor in units of 
$\Omega$. The 
values for superconducting YBa$_2$Cu$_3$O$_{7-\delta}$ are for microwave frequency 5.5 GHz and the 
values for normal materials are for dc.} \label{table:rho}
\begin{indented}
\item[]\begin{tabular}{@{}llll}
\br
        & 21 K & 75 K & 85 K \\ 
\mr
FIB-YBCO &  $1.7\times10^{-8}$ &  $1.7\times10^{-7}$ & 
                                                        $2.3\times10^{-6}$
                                     \protect\rule[0cm]{0pt}{3ex} \\
YBCO &  $7.2\times10^{-12}$ &  $2.55\times10^{-11}$ & 
                                                       $1.1\times10^{-6}$ \\
Copper &  $1\times10^{-10}$ & $2\times10^{-9}$ & 
                                                        $2\times10^{-9}$ \\   
\br
\end{tabular}
\end{indented}
\end{table}

It is remarkable that the resistivity (ie. the losses per unit volume) of the damaged YBa$_2$Cu$_3$O$_{7-\delta}$ 
is much larger than the undamaged YBa$_2$Cu$_3$O$_{7-\delta}$ and even larger than for copper. 
At 21 K, the damaged YBa$_2$Cu$_3$O$_{7-\delta}$ is still superconducting, but exhibits quite large microwave losses.
The resistivity 
at 75~K of the FIB-YBa$_2$Cu$_3$O$_{7-\delta}$ approaches the normal value at 85~K. Consistent with much of the 
damaged material being normal at 75~K and with the assumption that the extra microwave
loss is due to the microwave damage.

The physical damage caused by the FIB writing can be compared to the electrical damage. 
Doping due to Si ion implantation is
about 1 part in $10^5$ which is very small and we wouldn't expect this to have a large effect.
The resulting number of vacancies (about 1500 vacancies per Si ion from the TRIM calculation) 
is about 2\%  of the total YBa$_2$Cu$_3$O$_{7-\delta}$ atoms.
This damage results in the doubling of the normal resistivity and the lowering of $T_c$
by 5~K or about 6\%. The effects on the microwave loss are more dramatic. The microwave loss in the 
FIB damaged material at 21~K, was increased by a factor of 6000 over the loss in the undamaged YBa$_2$Cu$_3$O$_{7-\delta}$.
The increase in the loss at 75~K was by a factor of 16,000.
A small amount of physical damage has a large effect on normal resistivity and $T_c$
and a huge effect on microwave losses.

\section{Conclusions}

 We measured the microwave loss of a
resonator constructed with a film which had a 50~nm line written across
it by FIB irradiation. We did not observe evidence for Josephson effects
in this film. The microwave surface impedance measurements highlight the
variance of the power dependence seen in individual films. Massive
damage may be inflicted upon the film without altering the power
dependence even though the TE10 mode requires a high screening current
density in this region. However, mass creation of defects is observed to
increase the residual surface resistance of the film. These microwave power 
independent losses are not due to Josephson effects or the granular nature of the
thin film, but due to microscopic defects.

The as grown YBa$_2$Cu$_3$O$_{7-\delta}$ films start with a large concentration of
defects. We have seen that adding more defects leads to a large
increase in $R_s$. On the other hand, adding defects to the best
quality single crystals lowers $R_s$ (by decreasing the quasiparticle
scattering time). Perhaps there is an optimum concentration of
defects at a level between these two extremes. Noting the low
value of $R_s$  for good single crystals ($R_s\sim 10^{-5}~\Omega$ at about 5
GHz~\cite{hosseini}), there is still room
for improvement of YBa$_2$Cu$_3$O$_{7-\delta}$ thin films, if they can be grown with
fewer defects.

\ack We would like to thank L. E. Erickson, and G. Champion
who did the FIB writing and H.T. Tran who grew the YBa$_2$Cu$_3$O$_{7-\delta}$ films at NRC.

Two of the authors (AC and LC) would like to thank the Royal Society, EPSRC, and
National Physical Laboratory for their support.

\end{document}